\documentclass[prb,noshowpacs,aps,amssymb,showpacs,onecolumn,amsmath]{revtex4}

\newcommand{\beq}{\begin{equation}}
\newcommand{\eneq}{\end{equation}}
\usepackage{graphicx}
\input{epsf}

\begin{document}

\tolerance 10000


\title{Geometry of the 3-Qubit State, Entanglement and Division Algebras}
\author { Bogdan A. Bernevig}
\affiliation{ Department of Physics, Stanford University,
         Stanford, California 94305}
\author {Han-Dong Chen}
\affiliation{ Department of Applied Physics, Stanford University,
         Stanford, California 94305}
\begin{abstract}
\begin{center}

\parbox{13cm}{We present a generalization to $3$-qubits of the standard Bloch sphere representation for a single qubit
and of the 7-dimensional sphere representation for 2 qubits
presented in Mosseri {\it et al.}\cite{Mosseri2001}. The Hilbert space of the $3$-qubit system is
the $15$-dimensional sphere $S^{15}$, which allows for a natural
(last) Hopf fibration with $S^8$ as base and $S^7$ as fiber. A
striking feature is, as in the case of $1$ and $2$ qubits, that
the map is entanglement sensitive, and the two distinct ways of
un-entangling $3$ qubits are naturally related to the Hopf map. 
We define a quantity that measures the degree of entanglement of the 
$3$-qubit state.
Conjectures on the possibility to generalize the construction for higher qubit
states are also discussed.}

\end{center}
\end{abstract}

\maketitle

\section{Introduction}

Quantum mechanics exhibits its difference from classical physical
theories in many aspects. 
 A quintessential property of quantum
mechanics is quantum entanglement. Quantum entanglement rests at
 the center of the applications such as quantum information
and quantum computing. 
Maximally entangled EPR pair\cite{EPR} is an essential ingredient of teleportation\cite{Bennett1992}, dense coding\cite{Bennett1993}, and quantum key distribution\cite{{Ekertt1991},{Ekertt1992}}. The maximally entangled 3-qubit GHZ state\cite{GHZ} and the $m$-cat state are of cardinal importance to the applications such as cryptographic conferencing or superdense coding \cite{Bose1998}, quantum secret sharing or quantum information splitting\cite{Hillery1999}.
Due to the entanglement of the Hilbert space states, it is a highly non-trivial problem to understand the properties of multi-qubit states.
Recently, it has become clear\cite{Mosseri2001} that the properties of the first two simplest qubit states, the single qubit and the $2$-qubit state, are very deeply related to two very important mathematical objects, the first two Hopf fibrations $S^3\overset{S^1}{\longrightarrow}S^2$ and $S^7\overset{S^3}{\longrightarrow}S^4$. The global phase freedom of the single qubit state and the entanglement which appears for the first time starting with the $2$-qubit case have been proven to be deeply related to the Hopf fibrations. For an entangled $2$-qubit state, performing a transformation on the first qubit space induces a transformation on the space of the second qubit space. This feature is naturally captured by the nontrivial second Hopf fibration. The Hopf fibration can determine if the $2$-qubit state is entangled or separable\cite{Mosseri2001} and can also point to the degree of entanglement of a generic $2$-qubit pure state. Since obtaining a measure for the degree of entanglement is an essential issue of quantum computing, we believe it is extremely important if this method could be generalized to higher qubit states. Although attempts have been made towards describing the geometry of the 3-qubit states\cite{{Kus2001},{Carteret2000}}(Mosseri {\it et al.}\cite{Mosseri2001} briefly mentioned the generalization of their construction to include the $3$-qubit state), to our knowledge, no complete description is available.
 In this paper, we generalize the
discussion to the $3$-qubit state and the $3^{rd}$ Hopf fibration related to the last division algebra of the octonions. The
entanglement is understood in a geometrical way and a quantitative
measurement of entanglement is proposed. We describe the $3$-qubit Hilbert space as a nontrivial $S^7$ fibration over $S^8$. The entanglement quantity is proven to give the literature established values for the GHZ and W states.
The apparent failure of the algorithm for higher qubit states is also briefly discussed.
We would like to stimulate discussion and progress on the proper $n$-qubit generalization as the rewards obtained from such a generalization could prove enormous, possibly leading to a full classification of entanglement. We want to mention that, as it stands, our discussion is applicable to pure states only.

The paper is organized as follows. In Section \ref{section-single-qubit}
 we briefly recall
some well know facts about the $1$-qubit state, the Bloch sphere
representation and the close relation to the 1st Hopf fibration.
In section \ref{section-two-qubit}
 we present the recent results of Mosseri {\it et al.}\cite{Mosseri2001} which
relate the 2-qubit state to the second Hopf map ($S^7\overset{S^3}{\longrightarrow}S^4$). In
Section \ref{section-three-qubit} we begin the treatment of the 3-qubit state and
convincingly prove that it is related to the third and last Hopf
fibration thus clearly determining the geometry of the 3-qubit
state. We propose a quantity which can be used as a measure of
the entanglement of the 3-qubit state and comment on the
prospective generalizations to higher qubit states. Although not
strictly necessary, we use the language of the octonions, which
nicely simplifies notation and points to very interesting and deep
mathematical correspondences. In the appendix we give a brief
introduction to the octonions and the three Hopf maps which we
believe will be useful to a better understanding of the paper.

\section{Single Qubit, Bloch Sphere and $1^{st}$ Hopf Fibration}
\label{section-single-qubit}

The {\bf{pure}} 1-qubit state can be represented as a linear
combination of up and down spins:
\begin{equation}
|\Psi \rangle = \alpha_0 |0\rangle + \alpha_1 |1 \rangle , \;\;\;
\alpha_0, \alpha_1 \in \mathbb{C}, \;\;\; |\alpha_0|^2 + |\alpha_1|^2 =1,
\end{equation}
\noindent where we can parametrize
\begin{equation} \left(%
\begin{array}{c}
  \alpha_0 \\ ~\\
  \alpha_1 \\ 
\end{array}%
\right) =
\left(%
\begin{array}{c}
  \cos(\frac{\theta}{2})\exp(i\frac{\phi}{2} + i \frac{\chi}{2} ) \\ ~\\
  \sin(\frac{\theta}{2}) \exp(i\frac{\phi}{2} - i \frac{\chi}{2}) \\
\end{array}%
\right),  \;\;\;\;\;\; \theta \in [0, \pi], \;\;\; \phi \in
[0,2\pi], \;\;\; \chi \in [0, 2\pi].
\end{equation}
\noindent
 The Hilbert space of a single qubit with fixed norm unity
is the unit 3-dimensional sphere $S^3$. But since quantum mechanics is $U(1)$
projective, the projective Hilbert space is defined
up to a phase $\exp{(i\phi)}$. Therefore the
projective Hilbert space is $S^3/U(1)= S^3/S^1 = CP_1 = S^2$. This
property points to a map between the full Hilbert space $S^3$ and
the projective Hilbert space $S^2$, with the inverse map (fiber)
being an $S^1$. This map is the well known $1^{st}$ Hopf map,
$S^{3}\overset{S^{1}}{\longrightarrow}S^{2}$ which gives $S^3$ as an $S^1$ fibration over a base space $S^2$, the first in a series of maps that are deeply related to the structure of consistently defined number structures (division algebras, see Appendix). The map has the explicit form
\begin{subequations}
\begin{eqnarray}
  && h_1: \begin{array}{ccc}
    \mathbb{C} \otimes \mathbb{C} & \longrightarrow & \mathbb{C} \cup \{\infty\} \approx S^2 \\
    (\alpha_0, \alpha_1) & \longrightarrow & h_1 = \alpha_0 \alpha_1^{-1} \\
  \end{array}, \;\;\;\; |\alpha_0|^2 + |\alpha_1|^2 =1
 \\ 
 && h_2: \begin{array}{ccc}
    \mathbb{C} \cup \{\infty\} & \longrightarrow & S^2 \\
    h_1 & \longrightarrow & X_{i, \; (i=1,2,3)} \\
  \end{array}, \;\;\;\;~~ \sum_{i=1}^3 X_i^2 = 1 \;\; \\
&& h_2\circ h_1(\alpha_0, \alpha_1) =  
X_i = \langle \sigma_i
\rangle_{\Psi} = (\alpha_0^\star, \alpha_1^\star) \sigma_i \left(%
\begin{array}{c}
  \alpha_0 \\
  \alpha_1 \\
\end{array}%
\right),
\end{eqnarray}
\end{subequations}
\noindent where $\sigma_i(i=1,2,3)$ are the three Pauli matrices.
We can clearly see that the $X_i$'s are defined up to a $U(1)$
ambiguity in $\alpha_0, \alpha_1$. This map is very useful in
describing the density matrix for one qubit. The most general form
of this matrix is:
\begin{equation}
\rho= \frac{1}{2} (I + X_1 \sigma_1 + X_2 \sigma_2 + X_3 \sigma_3)
= \frac{1}{2} \left(%
\begin{array}{cc}
  1+X_3 & X_1 - i X_2 \\
  X_1 + iX_2 & 1-X_3 \\
\end{array}%
\right),
\end{equation}
\noindent with the constraint $\det \rho = 1- X_1^2 - X_2^2 -
X_3^2 \ge 0 $. For {\bf{pure}} qubit states, $\det \rho =0$. The
complete description of the single qubit Hilbert space and its essential
phase freedom can therefore be understood through the first Hopf fibration.
This fibration is nontrivial since  $S^3\neq S^1\otimes S^2$.
Physically, this means that it is impossible to consistently
ascribe a definite phase to each point on the Bloch sphere.

\section{Two Qubits, Entanglement and the $2^{nd}$ Hopf Fibration}
\label{section-two-qubit}

This section summarizes the results of Mosseri {\it et al}\cite{Mosseri2001}. A
{\bf{pure}} 2-qubit state reads:
\begin{subequations}
\begin{eqnarray}
&&|\Psi \rangle = \alpha_0 |00\rangle + \alpha_1 |01\rangle +
\beta_0 |10\rangle + \beta_1 |11 \rangle ,\\
&& \alpha_0,
\alpha_1, \beta_0, \beta_1 \in \mathbb{C}, \;\;\;\; |\alpha_0|^2 +
|\alpha_1|^2 + |\beta_0|^2 + |\beta_1|^2 =1.
\end{eqnarray}
\end{subequations}
\noindent The normalization condition means the Hilbert space of
the 2-qubit state  with fixed norm unity is a seven dimensional sphere $S^7$ and the projective Hilbert space is $S^7 / U(1) = CP_3$. The Hilbert space $\varepsilon$ is the tensorial product of single bit Hilbert spaces $\varepsilon_1 \otimes \varepsilon_2$.
In general, performing a transformation on the first qubit space
induces a transformation on the space of the second qubit space.
However, for the case in which $\alpha_0 \beta_1 =  \alpha_1
\beta_0$ one can independently transform the spaces of the two
single qubits. We then call the state non-entangled or separable.
To gain insight in the geometry and structure of the 2-qubit we
need to analyze the $S^7$ manifold of the Hilbert space . $S^7$
can be parametrized in many different ways as a product of
manifolds, but the most interesting parametrization\cite{Mosseri2001} is as an $S^3$fiber over an $S^4$. 
Notation can be greatly simplified by introducing a pair of quaternions:
\begin{subequations}
\begin{eqnarray}
q_1 &=& \alpha_0 + \alpha_1 i_2 = Re\alpha_0 + i_1 Im \alpha_0 + i_2
Re\alpha_1 + i_3 Im \alpha_1, \\
q_2 &=& \beta_0 + \beta_1 i_2  = Re \beta_0 + i_1 Im \beta_0 + i_2 Re
\beta_1 + i_3 Im \beta_1.
\end{eqnarray}
\end{subequations}
\noindent $i_1,i_2,i_3$ are square roots of $-1$ and form a basis for
the imaginary part of the quaternionic space ($\mathbb{Q} \sim \mathbb{R}^4, Im\mathbb{Q}\sim R^3$, see appendix).  Similar with the single qubit case, we can now define the following map

\begin{subequations}
\begin{eqnarray}
&&  h_1: \begin{array}{ccc}
    \mathbb{Q} \otimes \mathbb{Q} & \longrightarrow & \mathbb{Q} \cup \{\infty\} \approx S^4 \\
    (q_1, q_2) & \longrightarrow & h_1 = q_1 q_2^{-1} \\
  \end{array}, \;\;\; |q_1|^2 + |q_2|^2 =1 \\
 && h_2: \begin{array}{ccc}
    \mathbb{Q} \cup \{\infty\} & \longrightarrow & S^4 \\
    h_1 & \longrightarrow & X_{i, \;\; (i=1,...,5)} \\
  \end{array}, \;\;\;\; \sum_{i=1}^5 X_i^2  = 1 \;\;\;\; \\
 && h_2\circ h_1(q_1, q_2) =  X_i = \langle \sigma_i
\rangle_{\Psi} = (q_1^\star, q_2^\star) \sigma_i \left(%
\begin{array}{c}
  q_1 \\
  q_2 \\
\end{array}%
\right),
\end{eqnarray}
\end{subequations}
\noindent where 
\begin{eqnarray}
\sigma_1=\left(%
\begin{array}{cc}
  0 & 1 \\
  1 & 0 \\
\end{array}%
\right), \sigma_2=\left(%
\begin{array}{cc}
  0 & i_1 \\
  -i_1 & 0 \\
\end{array}%
\right), \sigma_3=\left(%
\begin{array}{cc}
  0 & i_2 \\
  -i_2 & 0 \\
\end{array}%
\right), \sigma_4=\left(%
\begin{array}{cc}
  0 & i_3 \\
  -i_3 & 0 \\
\end{array}%
\right), \sigma_5=\left(%
\begin{array}{cc}
  1 & 0 \\
  0 & -1 \\
\end{array}%
\right)
\end{eqnarray}
are a generalization of the Pauli matrices to Quternionic space.

The points $(q_1, q_2)$ and $(q_1 q, q_2 q)$ where
$q$ is a unit quaternion ($S^3$)  are mapped onto the same point
of the base space $S^4$ and therefore the map is a nontrivial
fibration $S^7\overset{S^3}{\longrightarrow}S^4$. This fibration
is entanglement sensitive\cite{Mosseri2001} in the sense that the
separable states defined by ${\alpha_0 \beta_1 =  \alpha_1 \beta_0}$ will be mapped onto the subset of pure complex
numbers in the Quaternion field, {\it i.e.}, 
\begin{eqnarray}
   X_3\biggl|_{\alpha_0 \beta_1 =  \alpha_1 \beta_0}=X_4\biggl|_{\alpha_0 \beta_1 =  \alpha_1 \beta_0}=0
~~~~or~~~
h_1(q_1, q_2)\biggl|_{\alpha_0 \beta_1 =  \alpha_1 \beta_0} \in
\mathbb{C}\subset \mathbb{Q}.    
\end{eqnarray}
It follows that the base space simplifies to a $S^2$
for non-entangled (separable) qubits. The partially traced density
matrix $\rho_1$ can be written as a functional of the variables in
the base space \cite{Mosseri2001}:
\begin{eqnarray}
    \rho_1&=&Tr_2 \rho=  \left[I_1\otimes(|0\rangle\langle0|+|1\rangle\langle1|)_2\right]
    |\Psi\rangle\langle \Psi|=\frac{1}{2}\left(
    \begin{array}{cc}
    1+X_5~& x_1-iX_2\\
    X_1+iX_2~&1-X_5
\end{array}
    \right).
\end{eqnarray}
This is the most general density matrix for a 1-qubit system. The
Bloch ball for one-qubit is then recovered from the 2-qubit
system by the partial trace. The determinant of $\rho_1$ is
\begin{eqnarray}
   \det \rho_1 = 1-X_1^2-X_2^2-X_5^2~=X_3^2+X_4^2.
\end{eqnarray}
$\det\rho_1=0$ for non-entangled qubits. Therefore, the density matrix
$\rho_1$ represents a pure state if $|\Psi\rangle$ is
non-entangled. Otherwise, $\rho_1$ represents a mixed state.
Mathematically, losing the information of the second qubit means 
integrating out or partial tracing the degree of freedom of the
second qubit. Then the resulted density matrix is only related to the
base space. It then follows naturally that the information of the
second qubit is stored in the fiber space while the information of
the first qubit and the correlation between these two qubits is
stored in the base space. In the non-entangled case, the $1^{st}$
Hopf map can be applied to the fiber $S^3$ (Hilbert space of the
second qubit) as described in the previous section. This would mod out the phase degree of the freedom. 
Finally, the $S^7$ fibration simplifies to $S^2 \otimes S^2$ for non-entangled
qubits, with one $S^2$ from the base and the other one from the
fiber. In addition, the quantity $X_3^2+X_4^2$ might be useful to
quantitatively measure the entanglement\cite{Mosseri2001}.

\section{Three Qubits, Entanglement and the $3^{rd}$ Hopf Fibration} 
\label{section-three-qubit}
It is interesting to see that the 1-qubit and
2-qubit systems are closely related to the first two Hopf
fibrations and the division algebras of the complex numbers and
the quaternions. This relation points to both
insightful comments on the geometry of the Hilbert space, and quantities which might describe entanglement. The 2-qubit system is the only system for which entanglement problem has so far been solved \cite{Bennett1996}
Multiple complications arise for higher qubit problems\cite{{Lewenstein2001},{Horodecki2000}}. In this section we go one step further to the first complicated qubit state, the 3-qubits. We show that its Hilbert
space geometry can be closely related to the geometry of the third
and last Hopf fibration and prove several insightful relations on
the entanglement of such state.

\subsection{The $3$-qubit Hilbert space. 2-qubit $\otimes$ 1-qubit entanglement}
\label{section-IV-A} 
The Hilbert space for the $3$-qubit is the tensor product of the
1-qubit Hilbert spaces $\varepsilon_1 \otimes \varepsilon_2 \otimes \varepsilon_3$ with a direct product basis: $\{ |000\rangle, |001\rangle, |010\rangle, |011\rangle, |100\rangle, |101\rangle,|110\rangle, |111\rangle \}$. A pure 3-qubit state reads:
\begin{subequations}
\begin{eqnarray}
  |\Psi \rangle = \alpha_0 |000\rangle + \alpha_1 |001\rangle +
\beta_0 |010 \rangle + \beta_1 |011 \rangle + \delta_0 |100\rangle
+ \delta_1 |101 \rangle + \gamma_0 |110 \rangle + \gamma_1
|111\rangle;\label{3-qubit} \\\nonumber \\
    \alpha_0, \alpha_1, \beta_0, \beta_1, \delta_0, \delta_1,\gamma_0,\gamma_1
\in \mathbb{C}, \;\;\; |\alpha_0|^2 + |\alpha_1|^2 + |\beta_0|^2 +
|\beta_1|^2 + |\delta_0|^2 + |\delta_1|^2 + |\gamma_0|^2+
|\gamma_1|^2 =1. \label{3-qubit-norm}
\end{eqnarray}
\end{subequations}
\noindent Differently from the case of 2-qubits, there are now
two ways in which the 3-qubit state can be separated. In the first
case, the 3-qubit case can be separated in the subspace of a
single qubit with basis $\{|0\rangle, |1 \rangle\} $ and the subspace
of 2-qubit $\{|00\rangle, |01\rangle, |10\rangle |11\rangle\}$:
\begin{subequations}
\begin{eqnarray}
  |\Psi \rangle = (a |0\rangle + b |1 \rangle)\otimes(c |00\rangle +
d|01\rangle + e|10\rangle + f |11 \rangle); \\ \nonumber\\
  a,b,c,d,e,f \in \mathbb{C}, \;\;\; (|a|^2+|b|^2)(|c|^2+|d|^2+|e|^2+|f|^2)=1.
\end{eqnarray}
\end{subequations}
\noindent In this scenario, we get the following relations:
\begin{equation}
\alpha_0 \gamma_1 = \delta_0 \beta_1 , \;\; 
\alpha_0 \gamma_0 = \delta_0 \beta_0 , \;\;
\alpha_0 \delta_1 = \delta_0 \alpha_1 , \;\;
\alpha_1 \gamma_1 = \delta_1 \beta_1 ,  \;\;
\alpha_1 \gamma_0 = \delta_1 \beta_0, \;\; 
\beta_0  \gamma_1 = \gamma_0 \beta_1.
\label{3-qubit-condition}
\end{equation}
\noindent Among these six conditions, only four are fundamental, from which the other two can be obtained. 

We can also go one step further and separate the $2$-qubit
subspace. In this case, the $3$-qubit state becomes fully separated
in the 3 1-qubit subspaces.
\begin{subequations}
\begin{eqnarray}
  |\Psi \rangle = (a |0\rangle + b |1 \rangle)\otimes(c |0\rangle + d |1 \rangle)
  \otimes(e |0\rangle + f |1 \rangle); \\ \nonumber\\
  a,b,c,d,e,f \in \mathbb{C}, \;\;\; (|a|^2+|b|^2)(|c|^2+|d|^2)(|e|^2+|f|^2)=1.   
\end{eqnarray}
\end{subequations}
\noindent The first step towards
separating the $3$-qubit space is the partial $1$-qubit $\otimes$ $2$-qubit separation.

The normalization condition (\ref{3-qubit-norm}) for the general $3$-qubit state 
identifies its Hilbert space with the 15 dimensional sphere
$S^{15}$. This manifold can be parametrized in many ways, but
considering the experience of the two previous sections and reminding ourselves of the existence of a third and last Hopf fibration $S^{15}\overset{S^7}{\longrightarrow} S^8$, it is tempting to see whether
it plays a role in the Hilbert space description.

\subsection{Octoninic representation of 3-qubit state and the third Hopf Fibration}

The most aesthetic way to introduce this fibration is with the use
of octonions instead of quaternions or complex numbers. Using
octonions introduces complications since they are not
only noncommutative (like the quaternions) but also
non-associative (see appendix). However, we feel that this
discomfort is compensated by the fact that the
mathematics becomes very compact and the connection with division algebra and the Cayley-Dickson construction (see appendix) becomes much clearer.

The construction of the two octonions from the complex
coefficients of the $3$-qubit state in Eq.(\ref{3-qubit}) proceeds as follows: we
first define 4 quaternions:
\begin{equation}
q_1 = (\alpha_0, \alpha_1) = \alpha_0 + \alpha_1 i_2, \;\; q_2
=(\beta_0, \beta_1)=\beta_0 + \beta_1^\star i_2, \;\; q_3 = (\delta_0,
\delta_1) = \delta_0 + \delta_1 i_2, \;\; q_4 = (\gamma_0,
\gamma_1) = \gamma_0 + \gamma_1^\star i_2.
\end{equation}
\noindent They satisfy the normalization $|q_1|^2+|q_2|^2 +|q_3|^2 + |q_4|^2 =1$. Out of these 4 quaternions, by the Cayley-Dickinson construction we can create two octonions belonging to the 8 dimensional octonionic space $\mathbb{O}\sim R^8$:
\begin{equation}
o_1 = (q_1, q_2) = q_1 + q_2 i_4, \;\;\;\; o_2 = (q_3, q_4) = q_3
+ q_4 i_4.
\end{equation}
\noindent The normalization condition now translates into $|o_1|^2 +|o_2|^2 =1 $, parametrizing an $S^{15}$. $i_1,i_2,i_3,i_4$ generates through multiplications 
$i_5,i_6,i_7$. These $7$ imaginary square roots of $-1$, along with the unity, close the octonionic multiplication table (see Appendix). The choice in the
definition of the four quaternions is specifically related to the
tensor-product nature of the 3-qubit Hilbert space. Had we made a
different choice for the 4 quaternions (2 octonions), we would
have induced an anisotropy on $S^{15}$, much in the same case as
in Mosseri {\it et al}\cite{Mosseri2001}. The Hopf map from $S^{15}$ to $S^8$ can again
be described as a map $h_1$ from $\mathbb{O} \otimes \mathbb{O}$
to $\mathbb{O} \cup \infty$ composed with an inverse stereographic
map $h_2$ from $\mathbb{O} \cup \infty$ to $S^8$:
\begin{subequations}
\begin{eqnarray}
  &&h_1: \begin{array}{ccc}
    \mathbb{O} \otimes \mathbb{O} & \longrightarrow & \mathbb{O} \cup \{\infty\} \approx S^8 \\
    (q_1, q_2) & \longrightarrow & h_1 = o_1 o_2^{-1} \\
  \end{array}, \;\;\; |o_1|^2 + |o_2|^2 =1
 \\
 && h_2: \begin{array}{ccc}
    \mathbb{O} \cup \{\infty\} & \longrightarrow & S^8 \\
    h_1 & \longrightarrow & X_{i, \;\; (i=1,...,9)} \\
  \end{array}, \;\;\;\; \sum_{i=1}^9 X_i^2  = 1 \;\;\;\; \\
& & h_2\circ h_1(o_1, o_2) =  X_i = \langle \sigma_i
\rangle_{\Psi} = (o_1^\star, o_2^\star) \sigma_i \left(%
\begin{array}{c}
  o_1 \\
  o_2 \\
\end{array}%
\right),
\end{eqnarray}
\end{subequations}
\noindent where 
\begin{eqnarray}
\sigma_1=\left(%
\begin{array}{cc}
  0 & 1 \\
  1 & 0 \\
\end{array}%
\right),~~\sigma_{2,3,4,5,6,7,8}=\left(%
\begin{array}{cc}
  0 & i_{1,2,3,4,5,6,7} \\
  -i_{1,2,3,4,5,6,7} & 0 \\
\end{array}%
\right),~~\sigma_9=\left(%
\begin{array}{cc}
  1 & 0 \\
  0 & -1 \\
\end{array}%
\right) 
\end{eqnarray}
are a generalization of the Pauli matrices to Octonionic
space. As in the case of previous Hopf maps, the fibration is not
trivial, as the space $S^8$ is not embedded in $S^{15}$. The fiber
is an seven dimensional sphere $S^7$, as can be seen by taking the
inverse map:
\begin{equation}
h_1^{-1} (y)= \left(%
\begin{array}{c}
  \{(yd,d) \; | \; d  \in \mathbb{O},\;\; |yd,d|=1 \}, \;\; x \ne \infty   \\
  \{(c,0) \; | \; c \in \mathbb{O}, \;\; |c|=1, \;\; x = \infty \\
\end{array}%
\right).
\end{equation}
\noindent We need to pause for a second and address an important
comment. Although in the case of quaternions which are only
non-commutative, it was clear that the map would have a unit
quaternion ($S^3$) as fiber, in the case of octonions, because of
their non-associativity, this is not automatically transparent.
However, the fact that the algebra is still alternative (see
appendix)(no other higher dimensional alternative algebra is
known) comes to our rescue and renders the fiber of the map be a
unit octonion $S^7$.

The first interesting feature of the fibration is revealed upon
explicit computation.
\begin{eqnarray}
    h_1(o_1,o_2) &=& o_1 o_2^{-1}
= \frac{C_1+C_2i_2+C_3i_4+C_4^\star i_6}{|\delta_0|^2 + |\delta_1|^2
+ |\gamma_0|^2 + |\gamma_1|^2}
\end{eqnarray}
with
\begin{subequations}
\begin{eqnarray}
C_1&=&\alpha_0 \delta_0^\star + \delta_1^\star \alpha_1 +
\gamma_0^\star \beta_0 + \beta_1 \gamma_1^\star \\
C_2 &=& \alpha_1
\delta_0 - \delta_1 \alpha_0 + (\beta_1 \gamma_0 - \gamma_1
\beta_0)^\star \\
C_3 &=& \beta_0 \delta_0 - \gamma_0 \alpha_0 +
(\alpha_1 \gamma_1 - \delta_1 \beta_1)^\star
\\
C_4 &=& \delta_1
\beta_0 - \alpha_1 \gamma_0 + (\beta_1 \delta_0 - \gamma_1
\alpha_0)^\star.
\end{eqnarray}
\end{subequations}
\noindent For the generic 3-qubit state, the $h_1$ map is octonionic in nature, as
we see above. However, for the case in which the 3-qubit state is
separable as a 1-qubit $\otimes$ 2-qubit, the $h_1$ maps into the subspace of pure complex numbers $\mathbb{C} \cup \infty$ in the octonionic field $\mathbb{O} \cup
\infty$:
\begin{equation}
h(o_1,o_2)\biggr|_{3=1\otimes 2} = \frac{C_1}{|\delta_0|^2 + |\delta_1|^2
+ |\gamma_0|^2 + |\gamma_1|^2}
 \;\;\; \in \;\; \mathbb{C}\cup \infty.
\end{equation}
\noindent We have just proved that the last Hopf map is
entanglement sensitive. In other words, by computing the value of
the map one can establish whether the 3-qubit state is entangled or is separable as an 1-qubit $\otimes$ 2-qubit state. We will come back to this later on as we define a quantity that characterizes the degree of entanglement and we will
see that the separated 2-qubit state lives on the fiber of the map
while the 1-qubit state lives on the base space of the map. The next step in
analyzing the geometry of the Hilbert space consists of an
analysis of the base space. For future reference, we here give the
expressions of the coordinates on the base space $S^8$: 
\begin{subequations}
\begin{eqnarray}
  X_1 & = & o_1 o_2^\star + o_2 o_1^\star \\
  X_2 & = & Re[i_1(o_1 o_2^\star - o_2 o_1^\star)] \\
  X_3 & = & Re[i_2(o_1 o_2^\star - o_2 o_1^\star)] \\
  X_4 & = & Re[i_3(o_1 o_2^\star - o_2 o_1^\star)] \\
  X_5 & = & Re[i_4(o_1 o_2^\star - o_2 o_1^\star)] \\
  X_6 & = & Re[i_5(o_1 o_2^\star - o_2 o_1^\star)] \\
  X_7 & = & Re[i_6(o_1 o_2^\star - o_2 o_1^\star)] \\
  X_8 & = & Re[i_7(o_1 o_2^\star - o_2 o_1^\star)] \\
  X_9 & = & o_1 o_1^\star - o_2 o_2^\star, 
\end{eqnarray}
\end{subequations}
\noindent where $o_1 o_2^\star - o_2 o_1^\star$ is purely
imaginary and $o_1 o_2^\star + o_2 o_1^\star $ and $ o_1 o_1^\star- o_2 o_2^\star$  are purely real. Their values are:
\begin{subequations}
\begin{eqnarray}
  o_1 o_1^\star - o_2 o_2^\star &=& \alpha_0 \alpha_0^\star 
  +\alpha_1 \alpha_1^\star+
  \beta_0 \beta_0^\star + \beta_1^\star \beta_1 - \delta_0 \delta_0^\star
  - \delta_1 \delta_1^\star - \gamma_0 \gamma_0^\star - \gamma_1^\star
  \gamma_1\\
  o_1 o_2^\star + o_2 o_1^\star  &=& \delta_0^\star \alpha_0 + \delta_1^\star \alpha_1 + \gamma_0^\star \beta_0 +
   \beta_1 \gamma_1^\star  +   \delta_0 \alpha_0^\star + \delta_1 \alpha_1^\star + \gamma_0 \beta_0^\star +
   \beta_1^\star \gamma_1 \\ 
  o_1 o_2^\star - o_2 o_1^\star&=& ( (a_0,a_1)\; , \; (b_0,b_1) ), \;\;\;\;\; a_0,a_1,b_0,b_1 \in \mathbb{C}
  \end{eqnarray}
\end{subequations}
with
\begin{subequations}
\begin{eqnarray}
  a_0 &=&  \delta_0^\star \alpha_0 + \delta_1^\star \alpha_1 + \gamma_0^\star \beta_0 +   \beta_1 \gamma_1^\star  -   \delta_0 \alpha_0^\star - \delta_1 \alpha_1^\star - \gamma_0 \beta_0^\star
   -  \beta_1^\star \gamma_1 \\
   a_1 &=&    2\alpha_1 \delta_0 - 2\delta_1 \alpha_0 +
  2 \beta_1^\star \gamma_0^\star - 2\gamma_1^\star \beta_0^\star \\
  b_0&=&2 \beta_0 \delta_0 - 2\gamma_0 \alpha_0
   + 2\alpha_1^\star \gamma_1^\star - 2\delta_1^\star \beta_1^\star \\
   b_1&=& 2\delta_1 \beta_0 - 2\alpha_1 \gamma_0 +
   2\beta_1^\star \delta_0^\star -2 \gamma_1^\star \alpha_0^\star
\end{eqnarray}
\end{subequations}
\noindent The 9 coordinates (subject to one constraint) of the
$S^8$ represent the generalization of the Bloch sphere
representation. For the case when the 3-qubit state is separable
as a 1-qubit $\otimes$ 2-qubit state, the map becomes purely
complex, as we have shown. In this case, 
\begin{eqnarray}
 o_1 o_2^\star - o_2 o_1^\star
 = \delta_0^\star \alpha_0 + \delta_1^\star \alpha_1 + \gamma_0^\star \beta_0 +
 \beta_1 \gamma_1^\star  -   \delta_0 \alpha_0^\star - \delta_1
\alpha_1^\star - \gamma_0 \beta_0^\star -  \beta_1^\star
\gamma_1,
\end{eqnarray}
which means $X_3=X_4=X_5=X_6=X_7=X_8=0$ which
means that only an $S^2$ ($X_1^2+X_2^2+X_9^2=1$) the base space
$S^8$ is used in the separable case. Therefore now things become
 clear: for a generic 3-qubit state, the Hilbert space is a
15 dimensional sphere $S^{15}$. This sphere admits many
parametrizations, the most famous of which is the third and last
Hopf map expressible as an $S^7$ fibration over $S^8$. As we have shown,
this fibration is entanglement sensitive, in the sense that it can
detect whether the 3-qubit state is separable as a product of a
1-qubit state and a 2-qubit state. Moreover, an analysis of
where the states are located points out that the 2-qubit state
occupies the fiber $S^7$ of the map while the single qubit state
occupies three ($X_1, X_2, X_9$) of the 9 coordinates on the base
space $S^8$. The rest of the coordinates somehow characterize the
degree of the entanglement between these two states, such that
they are zero -- as shown -- in the case when the 3-qubit states
is totally separable as a 1-qubit $\otimes$  2-qubit state.
Quantifying the degree of the entanglement will be our next priority. 
Since we have now established where the 2-qubit and the
single qubit states live, we now have a very similar picture to
the one developed by Mosseri {\it et al}\cite{Mosseri2001}. To obtain the fully
separable 3-qubit state into 3 1-qubit states, we first
separate it into a 1-qubit $\otimes$ 2-qubit state $S^2\otimes S^7$. We then focus on the fiber of the map, and use the
$2^{nd}$ Hopf fibration to separate it into an $S^2 \otimes S^3$ as
shown in the previous sections. We can then mod out the phase
degree of freedom by again particularizing to the fiber of the
second Hopf fibration and using the first Hopf fibration to mod
out an $S^1=U(1)$.

\subsection{Discussion}

Let's now obtain the the general expression for a state
$\Psi_O(\in S^{15})$ which is sent to $O$ by the map $h_1$.  The
inverse of the $3^{rd}$ Hopf map gives
\begin{eqnarray}
  \Psi_O=\left(\cos\Omega\exp(-\Theta {\bf T}/2)~o,~
      \sin\Omega\exp(\Theta {\bf T}/2)~o\right)
\end{eqnarray}
where $\cos\Theta=S(O')$, $\sin\Omega=X_1/S(O')$, $o$ is a unit octonion which spans the $S^7$ fiber and ${\bf T}$ is a unit pure imaginary octonion
\begin{eqnarray}
  {\bf T}=\frac{1}{\sin\Theta}\left(\sum_{m=1}^7{\bf V}_m(O')i_m\right).
\end{eqnarray}
Here $S(O')=(O'+(O')^*)/2$ and ${\bf V}(O')=(O'-(O')^*)/2$ are the scalar and vectorial parts of $O'\equiv O/|O|$.

\subsubsection{Separable states}
If the first qubit can be separated from the other two, $O$ is a complex number. Consequently, the state $\Psi_Q$ becomes
\begin{eqnarray}
  \Psi_O=\left(\cos\Omega\exp(-\Theta i/2)~o, \sin\Omega\exp(\Theta i/2)~o\right).
\end{eqnarray}
The $S^8$ base space reduces to $S^2$ sphere since
$X_3=X_4=X_5=X_6=X_7=X_8=0$. This $S^2$ sphere is exactly the Bloch sphere of the first qubit.

For the second and third qubits described by $o\in S^7$, we can define the coordinate system on fibre as
\begin{subequations}
\begin{eqnarray}
  |00\rangle_O&=&\left(\cos\Omega\exp(-i\Theta /2)|0\rangle_1
   +\sin\Omega\exp(i\Theta /2)|1\rangle_1\right)\otimes
   (|0\rangle_2\otimes|0\rangle_3)\\
  |01\rangle_O&=&\left(\cos\Omega\exp(-i\Theta /2)|0\rangle_1
   +\sin\Omega\exp(i\Theta /2)|1\rangle_1\right)\otimes
   (|0\rangle_2\otimes|1\rangle_3)\\
  |10\rangle_O&=&\left(\cos\Omega\exp(-i\Theta /2)|0\rangle_1
   +\sin\Omega\exp(i\Theta /2)|1\rangle_1\right)\otimes
   (|1\rangle_2\otimes|0\rangle_3)\\
  |11\rangle_O&=&\left(\cos\Omega\exp(-i\Theta /2)|0\rangle_1
   +\sin\Omega\exp(i\Theta /2)|1\rangle_1\right)\otimes
   (|1\rangle_2\otimes|1\rangle_3).
\end{eqnarray}
\end{subequations}

A generic state $\Psi_O$ in the $S^7$ fibre can be decomposed as
\begin{eqnarray}
    |\Psi_O\rangle=A_0|00\rangle_O+A_1|01\rangle_O
     + B_0|10\rangle_O+B_1|11\rangle_O
\end{eqnarray}
with $A_0,A_1,B_0,B_1\in \mathbb{C}$ and $|A_0|^2+|A_1|^2+|B_0|^2+|B_1|^2=1$. It's straightforward to see that the $3$-qubit system reduces to $1$-qubit $\otimes$ $2$-qubit. Now, we can fibrate the $S^7$ fiber space using the $2^{nd}$ Hopf map for this four-level $2$-qubit system. If this $2$-qubit is separable, the $S^7$ fiber space itself reduces to $S^2\otimes S^3$ with $S^3$ living on the fiber. Then we can again fibrate the $S^3$ to mod out the global phase.
   Consequently, if it is fully separable, the $3$-qubit reduces to $S^2\otimes S^2\otimes S^2$ with the first, second and third qubits living in the base space of the $S^{15}$ fibration, the base space of the $S^7$ fibration of the fibre and the fibre of $S^7$ fibration of the fibre, respectively.
%
\begin{figure}[h]
   \includegraphics[scale=0.35]{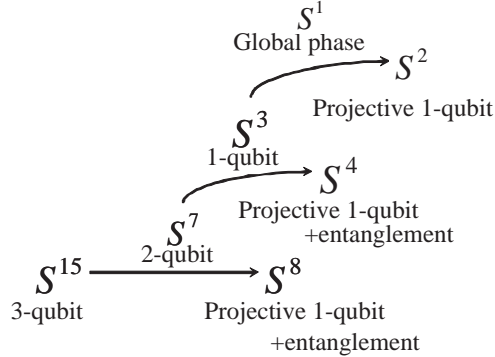}
   \caption{The iteration of Hopf fibration: $3$-qubit $\longrightarrow$ $1$-qubit$ \otimes$ $2$-qubit $\longrightarrow$ $1$-qubit $\otimes$ $1$-qubit $\otimes$ $1$-qubit.}
\end{figure}

\subsubsection{Entangled states}
Now, let us turn to the maximally entangled states(M.E.S.). They corresponding to the vector $C_2 i_2+C_3 i_3+C_4 i_6$ have maximal norm. For a M.E.S., $|\Psi_O\rangle$ reads
\begin{eqnarray}
  |\Psi_O\rangle = \frac{1}{\sqrt{2}}
  \left(
  \exp\left(-\pi\frac{C_2 i_2+C_3 i_4+ C_4^\star i_6}{2}\right)~o,~~
  \exp\left(\pi\frac{C_2 i_2+C_3 i_4+C_4^\star i_6}{2}\right)~o
  \right).\label{MES}
\end{eqnarray}
The M.E.S. expands a $5$-dimensional sphere $S^5$. For $C_4=\pm\frac{1}{2}$ and $o=(1\pm i_6)/\sqrt{2}$, the standard GHZ state is obtained from Eq.(\ref{MES}).

For any $\Omega.M.E.S$, $|\Psi_O\rangle$ can be written as
\begin{eqnarray}
  |\Psi_O\rangle =  \left(\cos\Omega
  \exp\left(-\frac{\pi}{4}\frac{C_2 i_2+C_3 i_4+ C_4^\star i_6}
  {|C_2 i_2+C_3 i_3+ C_4^\star i_6|}\right)~o,~~
  \sin\Omega\exp\left(\frac{\pi}{4}\frac{C_2 i_2+C_3 i_4+C_4^\star i_6}
  {|C_2 i_2+C_3 i_3+ C_4^\star i_6|}\right)~o
  \right)
\end{eqnarray}

From Eq.(\ref{entanglement-def}), one see the fact that the base space contains the information of the first qubit and the information of the correlation between it and the other two qubits while the fiber only contains the information of the second and third qubits only. We can utilize this observation to generalize the Bloch sphere representation. The Hopf map clearly suggests to split the representation into a product of base and fiber subspaces. For the base space $S^8$, we propose to only keep three coordinates:
\begin{eqnarray}
  (X_1,X_2,X_9)=(\langle\sigma_x\otimes I_{2-qubit}\rangle,
  \langle\sigma_y\otimes I_{2-qubit}\rangle,
  \langle\sigma_z\otimes I_{2-qubit}\rangle ).
\end{eqnarray}
All states are then mapped onto a ball $B^3$ of radius $1$ described by $0\leq X_1^2+X_2^2+X_9^2\leq 1$ . The set of separable states are mapped onto the $S^2$ boundary as discussed previously. The center of the ball corresponds to M.E.S. The concentric spherical shells correspond to the set of states with the same entanglement as defined in Eq.(\ref{entanglement-def}).

\subsubsection{Angle description of entanglement}
For a generic $3$-qubit state given by Eq.(\ref{3-qubit}),
we can decompose it as
\begin{eqnarray}
  |\Psi\rangle=|0\rangle|\Psi_0\rangle+  |1\rangle|\Psi_1\rangle
  \label{decomposition}
\end{eqnarray}
with
\begin{subequations}
\begin{eqnarray}
  |\Psi_0\rangle &=&  A_0 |00\rangle + A_1 |01\rangle +
B_0 |10 \rangle + B_1 |11 \rangle \\
   |\Psi_1\rangle &=&  A'_0 |00\rangle
+ A'_1 |01 \rangle + B'_0 |10 \rangle + B'_1
|11\rangle.
\end{eqnarray}
\end{subequations}
Geometrically, we can imagine that $|\Psi_0\rangle$ ($|\Psi_1\rangle$) lives on the north (south) pole of the one-qubit Bloch sphere. After parallel transporting the vector $|\Psi_1\rangle$ from the south pole to the north pole, we can define an angle to quantify the difference between $|\Psi_0\rangle$ and $|\Psi_1\rangle$. If these two vectors are pointing in the same direction, {\it i.e.},
\begin{eqnarray}
    |\Psi_0\rangle=C|\Psi_1\rangle~~~~~(C \in\mathbb{C}),\label{1x2-condition}
\end{eqnarray}
the first qubit can be separated from the other two and the $3$-qubit state $|\Psi\rangle$ reduces to a $1$-qubit $\otimes$ $2$-qubit state. We then can iterate this decomposition for the $2$-qubit state $|\Psi_0\rangle$.

\begin{figure}[h]
  \includegraphics[scale=0.25]{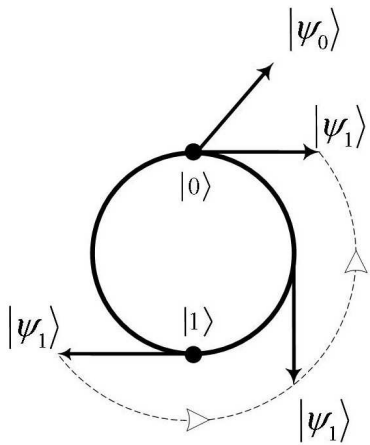}
  \caption{The graphical representation of the decomposition given by Eq.(\ref{decomposition}). After parallel transporting the vector $|\Psi_1\rangle$ from the south pole to the north pole, we can define an angle to quantify the difference between $|\Psi_0\rangle$ and $|\Psi_1\rangle$. This angle could be used to quantify the entanglement of the state $|\Psi\rangle$.}
  \label{Fig-entanglement}
\end{figure}

\noindent The condition (\ref{1x2-condition}) leads to the same conditions as given in Eq.(\ref{3-qubit-condition}).
A natural definition of the entanglement is then given by
\begin{eqnarray}
  E=A\sum_{b,c,b'c'} \biggl[|t_{0bc}t_{1b'c'}-t_{0b'c'}t_{1bc}|^2+
  |t_{b0c}t_{b'1c'}-t_{b'0c'}t_{b1c}|^2+|t_{b0c}t_{b'1c'}-t_{b'0c'}t_{b1c}|^2
  \biggr]\label{entanglement-angle}
\end{eqnarray}
where $A$ is a proper normalization factor and $t_{abc}\equiv\langle abc|\Psi\rangle$. The generalization of this definition is straightforward. This definition is exactly the same as the one discussed by Meyer {\it et al.}\cite{Meyer2002}.

\subsubsection{Quantifying Entanglement}

Based on our discussion above, we are now in position to propose a
quantity that quantifies the degree of entanglement of a 3-qubit
state. As our discussion so far suggests, we need to probe for the
entanglement of 3-qubits in a 1-qubit $\otimes$ 2-qubit state.
(Subsequently, we can particularize to the fiber of the third hopf
map and classify the degree of entanglement of the 2-qubit state.)
We therefore partially trace two qubits to obtain the partially
traced matrix $\rho_1$:
\beq \rho_1 = \frac{1}{2} \left(%
\begin{array}{cc}
  1+X_9 & X_1 - iX_2 \\
  X_1+iX_2 & 1-X_9 \\
\end{array}%
\right) \eneq 
\noindent Usually, for generic 3-qubit states, $\det\rho_1 > 0$. However, in the case of 2-qubit $\otimes$ 1qubit
entanglement, the determinant of the matrix vanishes, and
therefore $X_3=X_4=X_5=X_6=X_7=X_8=0$. It seems obvious then that
we could use the quantity 
\begin{eqnarray}
 E=X_3^2+ X_4^2 +X_5^2 +X_6^2 + X_7^2 +X_8^2 = 1 - X_1^2 -X_2^2 -X_9^2   
 \label{entanglement-def}
\end{eqnarray}
to quantify entanglement. Small values of $E$ means high degree of separability of 1-qubit and 2-qubit Hilbert spaces in the 3-qubit state and viceversa. Notice that this quantity $E$ only measures the entanglement between the first qubit and the $2$-qubit system of second and third qubit. Similarly, we can construct the second or third qubit into base space to get two different constructions of the Hopf fibration. A more reasonable definition of the measurement of the entanglement will be the average of the quantity given in Eq.(\ref{entanglement-def}) over all possible constructions.

Let us now test these assumptions on two well known states, GHZ
and W states of the 3-qubit problem. The generalized GHZ states
read:
\begin{equation}
|GHZ \rangle|_{generalized} =  \alpha_0 |000\rangle + \gamma_1
|111 \rangle \;\;\; \Longrightarrow \;\;\;
\begin{array}{c}
  X_5 = 2Re \gamma_1 Im \alpha_0 + 2Im \gamma_1 Re \alpha_0  \\
  X_6= 2Re \gamma_1 Re \alpha_0 - 2Im \gamma_1 Im \alpha_0 \\
  X_9  =  |\alpha_0|^2 - |\gamma_1|^2 \;\;\;\;\;\;\;\;\;\;\;\;\;\;\;\;\;\;\;\;\;\;\; \\
\end{array}
\end{equation}
\noindent the other $X$'s being zero, and with a degree of
entanglement $E= 1 - |\alpha_0|^2 + |\gamma_1|^2$. For the
{\bf{pure}} GHZ state $\alpha=\gamma_1=1/\sqrt{2}$ and therefore $E=1$,
meaning that the GHZ state is a maximally entangled state of the 3
qubit system, consistent with well-known result.

The generalized W state reads:
\begin{equation}
|W\rangle_{generalized} =\delta_0|100\rangle + \beta_0|010 \rangle + \alpha_1|001 \rangle
 \;\;\; \Longrightarrow \;\;\;\begin{array}{c}
  X_3 = 2 Re \alpha_1 Re \delta_0 - 2Im \alpha_1 Im \delta_0 \\
  X_4 = 2 Im \alpha_1 Re \delta_0 + 2Re \alpha_1 Im \delta_0 \\
  X_5 = 2 Re \beta_0 Re \delta_0 - 2Im \beta_0 Im \delta_0 \\
  X_8 = 2 Im \beta_0 Re \delta_0 + 2Re \beta_0 Im \delta_0 \\
  X_9= |\alpha_1|^2 + |\beta_0|^2 - |\delta_0|^2 \;\;\;\;\;\;\;\;\;\;\; \\
\end{array}
\end{equation}
\noindent For the W state, $X_3 = X_5 =2  X_9 = \frac{2}{3}$ and
the degree of entanglement is $E= X_3^2+X_5^2 = 8/9$, consistent with
the literature\cite{Meyer2002}.

\subsubsection{Conjecture}
A natural question is whether this construction is generalizable
to systems with more than 3 qubits. One can imagine expanding the
same formalism by always adding another square root of unity and
forming the next algebra. Although this is possible via the
Cayley-Dickenson formalism (see Appendix), the algebras formed in
this way are not alternative, and cannot be written as fibrations
of spheres over sphere base spaces. The Hopf construction stops at
octonions. However, the subsequent algebras, although not
division, are nicely normed, which means that they have an
inverse. So in principle the type of map that we give in this paper is
possible. However, the map would fail in the
following sense: it would be possible to map non-zero points into
zeros in the base space, fact which is not possible in the maps
using division algebra numbers. This is just a restatement of the
fact that further algebras would have zero divisors.

However, the Cayley-Dickson construction, as well as the fact that
the number of dimensions of the algebras created by this
construction is identical to the number of dimensions of the qubit
spaces, hint at some deeper connection between the Cayley
construction and qubit states. Interestingly, this construction might be very related to the hyperdeterminant construction of Miyake and Wadati\cite{miyake2003}. Out definition of the entanglement $E$ in Eq.(\ref{entanglement-angle}) is very similar to the hyperderterminant construction. It would be interesting to investigate this correspondence for higher qubit states. 
The non-existence of Hopf maps for
higher than 3 qubits seems to tell us that the 1-qubit, 2-qubit,
and 3-qubit states are, in some sense, more special than higher
qubit states. However, the richness of information that we are
able to procure with the identification presented in this paper
and in the paper by Mosseri and Dandoloff seems to make further
investigation in this field worthy.

\section{Conclusions}
\label{Conclusion}
In this paper we analyze the 3-qubit state. We give a full description of the 3-qubit Hilbert space by relating it to the third and last Hopf fibration. We prove that this fibration is entanglement sensitive, that is, it can detect whether the 3-qubit state is separable or entangled. Moreover, we show that one can define a quantity to describe the entanglement of the 3-qubit state and the possibility of it being separable as a 1-qubit$\otimes$ 2-qubit state. Our results, cumulated to the results of Mosseri, show that non-trivial fibrations are a very useful tool in describing many-qubit states and their entanglement.

\section{Acknowledgements}
This paper was the result of a suggestion by S.C. Zhang, for which we are deeply grateful. We also acknowledge private communications with Remy Mosseri, for which we are deeply grateful. The authors would like to thank G. Chapline, C.H.Chern, T. Cuk, J. Franklin,  R.B. Laughlin, D. Santiago, T.-C. Wei, C.J. Wu, J. T. Yard and G. Zeltzer for valuable discussions. This work is supported by the NSF under grant numbers DMR-9814289 and 2FEV602, and the US Department of Energy, Office of Basic Energy Sciences under
contract DE-AC03-76SF00515. The authors also acknowledge support from the Stanford
Graduate Fellowship Program. 

\appendix
\section{Octonions and the last Division Algebra}
An extensive review of octonions and division algebras is provided
by Baez\cite{baez}. Real and Complex numbers are used by physicists
daily. Although real numbers are in a sense `nicer' than complex
numbers because the conjugate of a real number is itself, complex
numbers bring about new and powerful properties and structure.
However, they are only the first two kind of numbers in a set of
four possible structures. In a far-reaching and very deep
argument, it has been proved that there are only four division
algebras, in other words, there are only 4 vector spaces $A$
equipped with a bilinear map $m: \; A \times A \rightarrow A$
called multiplication, and with a non-zero element called unit
such that $m(1,a) = m(a,1) = a$ (these properties form an algebra)
and given $a,b \in A$ with $ab=0$ then either $a=0$ or $b=0$ (no
zero divisors - property defining the division algebra). The real
and the complexes ($\mathbb{R}, \; \mathbb{C}$) form the first two
division algebras. The third and fourth division algebra are the
quaternions and the octonions ($\mathbb{Q}, \; \mathbb{O}$). The
Cayley-Dickson constructions provides a construction of the elements in
$\mathbb{R}, \; \mathbb{C}, \; \mathbb{Q}, \; \mathbb{O}$ which
makes apparent the fact that each one fits nicely in the next. The
complex numbers can be considered as a pair of real numbers
$(a,b)$; then addition can be performed component-wise whereas the
multiplication rule is:
\begin{equation}
(a,b)(c,d) = (ac - db, ad + cb) = ac-db + (ad+cb)i.
\end{equation}
\noindent We can define the quaternions in a similar way: a
quaternion is a pair of complex numbers $(a,b)$, with the complex
conjugation and the multiplication laws being:
\begin{equation}
(a,b)^\star = (a^\star, -b), \;\;\;\;\; (a,b)(c,d) = (ac - d^\star
b,bc^\star + da) = ac - d^\star b + (bc^\star + da) i_2.
\end{equation}
\noindent The quaternions are non-commutative and upon expansion,
can be written as $q =  Re a + i_1 Im a + i_2 Re b + i_3 Im b$,
$i_1^2 = i_2^2 = i_3^2 = i_1i_2i_3=-1$. We can go one step further and
build an octonion from a pair of quaternions $(q_1, q_2)$, with
the multiplication and conjugation laws the same as before. The
octonions are non-associative, as well as non-commutative. They
are the biggest division algebra. If one continues the
Cayley-Dickson construction further, by taking a pair of
octonions, one discovers that the division property is lost, that
is, the new numbers have zero divisors. The division algebras,
including the non-associative octonions have the essential property
that they are alternative, in other words:
\begin{equation}
\forall \;\; a,b \in \mathbb{R}, \mathbb{C}, \mathbb{Q},\mathbb{O}
\;\; \Longrightarrow \;\; (aa)b = (aab), \;\; (ab)a = a(ba), \;\;
(ba)a = b(aa).
\end{equation}
\noindent Octonions can be presented in the
double quaternion format but also, equivalently, in expanded
format with $i_1, i_2, i_3, i_4, i_6, i_7$ as imaginary units
(square roots of $-1$):
\begin{equation}
o= x_0 + \sum_{\alpha} x_m i_m, \;\;\;\; x_{0,...,7} \in
\mathbb{R}, \;\;\;\;\; i_{1}^2 = ... = i_7^2 = -1,
\end{equation}
\noindent which can also be described in terms of quaternions and
complex numbers as $o = \{(x_0 + x_1 i_1) + (x_2 + x_3 i_1)i_2\} +
\{(x_4 + x_7 i_1) + (x_6 - x_5 i_1)i_2\} i_4 $. The multiplication
table can be given in terms of the cycles:
\begin{equation}
(123), \;\; (246), \;\; (435), \;\; (367), \;\; (651), \;\; (572),
\;\; (714)
\end{equation}
\noindent which read, for example $i_7 i_1 = i_4$, etc. The
conjugate and inverse of an octonion $o$ is:
\begin{equation}
\bar{o} = x_0 - \sum_{m} x_m i_m, \;\;\; o^{-1} =
\frac{\bar{o}}{|o|^2}.
\end{equation}
\noindent Another way in which an octonion $o$ can be written is
as a scalar $S(o)$ part and a vectorial $V(o)$ part:
\begin{equation}
S(o) = \frac{1}{2} (o+\bar{o}) = x_0, \;\; V(o) = \frac{1}{2}
(o-\bar{o}) = \sum_{m=1}^7 V_m(o) i_m =
\sum_{m=1}^7 x_i i_m
\end{equation}
\noindent An octonion $o$ can also be written in exponential form:
\begin{equation}
o = |o| \exp (\theta I), \;\; \theta = \arccos\left(\frac{S(o)}{|o|}\right),
\;\; I = \frac{V(o)}{|V(o)|}
\end{equation}
\noindent As presented in the body of the paper, the third hopf
map is nicely presented in terms of octonions:
\begin{subequations}
\begin{eqnarray}
  &&h_1: \begin{array}{ccc}
    \mathbb{O} \otimes \mathbb{O} & \longrightarrow & \mathbb{O} \cup \{\infty\} \approx S^8 \\
    (q_1, q_2) & \longrightarrow & h_1 = o_1 o_2^{-1} \\
  \end{array}, \;\;\; |o_1|^2 + |o_2|^2 =1
 \\
 && h_2: \begin{array}{ccc}
    \mathbb{O} \cup \{\infty\} & \longrightarrow & S^8 \\
    h_1 & \longrightarrow & X_{i, \;\; (i=1,...,9)} \\
  \end{array}, \;\;\;\; \sum_{i=1}^9 X_i^2  = 1 \;\;\;\; \\
& & h_2\circ h_1(o_1, o_2) =  X_i = \langle \sigma_i
\rangle_{\Psi} = (o_1^\star, o_2^\star) \sigma_i \left(%
\begin{array}{c}
  o_1 \\
  o_2 \\
\end{array}%
\right),
\end{eqnarray}
\end{subequations}
\noindent where
\begin{eqnarray}
\sigma_1=\left(%
\begin{array}{cc}
  0 & 1 \\
  1 & 0 \\
\end{array}%
\right), \sigma_{2,3,4,5,6,7,8}=\left(%
\begin{array}{cc}
  0 & i_{1,2,3,4,5,6,7} \\
  -i_{1,2,3,4,5,6,7} & 0 \\
\end{array}%
\right), \sigma_9=\left(%
\begin{array}{cc}
  1 & 0 \\
  0 & -1 \\
\end{array}%
\right)
\end{eqnarray}
  are a generalization of the Pauli matrices to Quternionic
space. The form of the map is identical to the form of the map
presented in Mosseri {\it et al}\cite{Mosseri2001}. 
However, proving that $S^7$ is the
fiber of this map turns out to be non-trivial, since as opposed to
the quaternionic case of the second Hopf map, we lose the
associativity property and therefore $(o_1 o_2) o_3 \ne o_1(o_2o_3)$ for $o_1,o_2,o_3\in\mathbb{O}$. However, after some explicit calculations one can find out
that the essential property is that the algebra be alternative.
Alternativity holding, one can prove the following: $o_1^{-1} (o_1o_2) = (o_1^{-1} o_1) o_2$ and therefore inverse map is:
\begin{equation}
h_1^{-1} (y)= \left(%
\begin{array}{c}
  \{(yd,d) \; | \; d  \in \mathbb{O},\;\; |yd,d|=1 \}, \;\; x \ne \infty   \\
  \{(c,0) \; | \; c \in \mathbb{O}, \;\; |c|=1, \;\; x = \infty \\
\end{array}%
\right)
\end{equation}
\noindent


\begin{thebibliography}{10}
    \bibitem{Mosseri2001}
    Remy Mosseri and Rossen Dandoloff, J. Phys. A, {\bf 34} 10243 (2001).
     
     \bibitem{EPR}
    A. Einstein, B. Podolsky and N. Rosen, Phys. Rev. {\bf 47}, 777 (1935).
     
     \bibitem{Bennett1993}
     C.H. Bennett, G. Brassard, C. Crepeau, R. Joza, A. Peres, and W.K. Wootters,     Phys. Rev. Lett. {\bf 70}, 1895 (1993). 

    \bibitem{Bennett1992}
    C.H. Bennett and S. Wiesner, Phys. Rev. Lett. {\bf 69}, 2881 (1992). 

    \bibitem{Ekertt1991}
    A.K. Ekert, Phys. Rev. Lett. {\bf 67}, 661 (1991). 

    \bibitem{Ekertt1992}
    A.K. Ekert, J.G. Rarity, P.R. Tapster, and G.M. Palma, Phys. Rev. Lett. {\bf 69}, 1293 (1992). 
    
    \bibitem{GHZ}
    D. M. Greenberger, M. A. Horne and A. Zeilinger, in {\it Bell's Theorem, Quantum Theory and Conceptions of the Universe}, edited by M. Kafatos (Kluwer Academic, Dordrecht, 1989), p.69.
    
    \bibitem{Bose1998}
    S. Bose, V. Vedral, and P.L. Knight, Phys. Rev. A {\bf 57}, 822 (1998). 

    \bibitem{Hillery1999}
    M. Hillery, V. Buzek, and A. Berthiaume, Phys. Rev. A {\bf 59}, 1829 (1999). 
    
    \bibitem{Kus2001}
    M. Kus and K. Zyczkowski, Phys. Rev. A, {\bf 59} 032307 (1991).
    \bibitem{Carteret2000}
    H.A.Carteret and A. Sudbery, J. Phys. A, {\bf 33} 4981 (2000).

    \bibitem{Bennett1996}
    C.H.Bennet, {\it et al}, Phys. Rev. A, {\bf 53}, 2046 (1996);
    C.H.Bennet, {\it et al}, Phys. Rev. A, {\bf 54}, 3824 (1996);
    V. Vedral {\it et al}, Phys. Rev. Lett., {\bf 78}, 2275 (1997);
    V. Vedral {\it et al}, Phys. Rev. A, {\bf 56}, 4452 (1997);
    V. Vedral {\it et al}, Phys. Rev. A, {\bf 57}, 1619 (1998);
    M. Horodecki, {\it et al}, Phys. Rev. lett., {\bf 80}, 5239 (1998);
    E.M.Rains, {\it et al}, Phys. Rev. A, {\bf 60}, 179 (1999);
    S. Hill and W.K. Wootters, Phys. Rev. Lett., {\bf 78}, 5022 (1997);
    W.K.Wootters, Phys. Rev. Lett., {\bf 80}, 2245 (1998);
    A. F. Abouraddy, {\it et al}, Phys. Rev. A, {\bf 64} 050101 (2001).
            
    \bibitem{Lewenstein2001}
    M. Lewenstein, {\it et al}, J. Mod. Opt., {\bf 47} 2481 (2001).

    \bibitem{Horodecki2000}
    M.  Horodecki, {\it et al}, e-print quant-ph/0006071.

    \bibitem{Meyer2002}
    D. A. Meyer and Nolan R. Wallach, J. Math Phys., {\bf 43} 4273 (2002).
        
     \bibitem{baez}
     J. C. Baez, Bull. Amer. Math. Soc. {\bf 39} 145 (2002).
     
     \bibitem{miyake2003}
     A. Miyake, Phys. Rev. A {\bf 67}, 012108 (2003). A.Miyake and M. Wadati, quant-ph/0212146.
     
\end{thebibliography}

\end{document}